\documentclass[preprint,aps,prd,amsmath,amssymb]{mystyle}

\usepackage{graphicx}

\def\bfi{\begin{figure}}
\def\efi{\end{figure}}
\def\bc{\begin{center}}
\def\ec{\end{center}}
\def\be{\begin{equation}}
\def\ee{\end{equation}}
\def\bitz{\begin{itemize}}
\def\eitz{\end{itemize}}
\def\benum{\begin{enumerate}}
\def\eenum{\end{enumerate}}

\begin{document}
\begin{frontmatter}

\title{Characterization of large area photomutipliers under low magnetic fields: design and performances of the magnetic shielding for the Double Chooz neutrino experiment}

\author{E. Calvo},
\author{M. Cerrada},
\author{C. Fern\'andez-Bedoya},
\author{I. Gil-Botella},
\author{C. Palomares},
\author{I. Rodr{\'i}guez},
\author{F. Toral},
\author{A. Verdugo}

\address{Centro de Investigaciones Energ{\'e}ticas, Medioambientales y Tecnol{\'o}gicas (CIEMAT), Av. Complutense 22, 28040 Madrid, Spain}

\begin{abstract}
This paper describes the characterization studies under low magnetic fields of the Hamamatsu R7081 photomultipliers that are being used in the Double Chooz experiment. The design and performances of the magnetic shielding that has been developed for these photomultipliers are also reported.
\end{abstract}

\end{frontmatter}

\section{Introduction}
\label{intro}

The main goal of the Double Chooz experiment is to measure the mixing angle $\theta_{13}$ by searching for the disappearance of electron antineutrinos produced in the CHOOZ nuclear power station in Ardennes (France). Double Chooz will be the first neutrino experiment with enough sensitivity to improve the current upper limit~\cite{clim} in almost one order of magnitude in case no oscillation was observed.

In order to eliminate the main systematic error source related to the neutrino flux uncertainty, Double Chooz will use two identical neutrino detectors. The first one (near detector) will be located at 400 m from the two reactors to measure with good precision the flux and the energy spectrum of antineutrinos.
The second one (far detector) will search for a deficit of electron antineutrinos and will be installed at 1.05 km from the reactor cores, in the hall built for the CHOOZ experiment~\cite{CHOOZ}, at the end of the nineties. 
The Double Chooz detectors (Fig.~\ref{fig-DC}) are composed by concentric cylindrical volumes filled with liquids of similar density but different optical properties.
The innermost volumes conform the target and contain liquid scintillator. Surrounding these active volumes there is a non-scintillating buffer inside a stainless steel tank, to reduce the events from accidental background. A total of 390 photomultiplier tubes (PMT) of 10 in. diameter (Hamamatsu R7081) are distributed on the inside of the stainless tank to collect the scintillation light. A detailed description of the Double Chooz detector is given in Ref.~\cite{DC}.

Reducing background levels is a key factor to achieve the sensitivity goals in this type of experiments. The Double Chooz detector, in addition to its underground location and the veto systems, is shielded by 15 cm iron bars to prevent the rock radioactivity from reaching the scintillator.  
The photomultiplier tube response may be affected by the Earth's magnetic field and an additional non-uniform contribution from the iron shield. The iron bars have been demagnetized in order to prevent the magnetic field being higher than 1 G in the PMTs region.

Although most photomultiplier tubes are affected by the presence of magnetic fields, large area PMTs are particularly vulnerable due to the long trajectories that the photoelectrons emitted in the photocathode have to travel until they reach the first dynode.
PMTs with a large photocathode area have been extensively used in the last years in high energy physics for neutrino detection experiments in order to cover very large detection areas with the minimum number of channels.
Examples of this are the 20 in. R1449 Hamamatsu PMTs used in the Super-Kamiokande experiment, the 8 in. ETL 9351 in Borexino or the 10 in. R7081 of Antares~\cite{exp}. In these experiments the PMTs have to deal with the uniform Earth's magnetic field while the Double Chooz PMTs, as previously mentioned, will be immersed in a non-uniform and unknown magnetic field due to the presence of the iron shield.
In addition, this field could be different for near and far detectors.
The aim of the work reported in this paper is to determine the effect of magnetic fields up to 1 G on the performances of the Hamamatsu R7081 PMT and to find a solution for shielding these PMTs against the magnetic field.

 \bfi
\includegraphics[height=7.cm,width=9.cm]{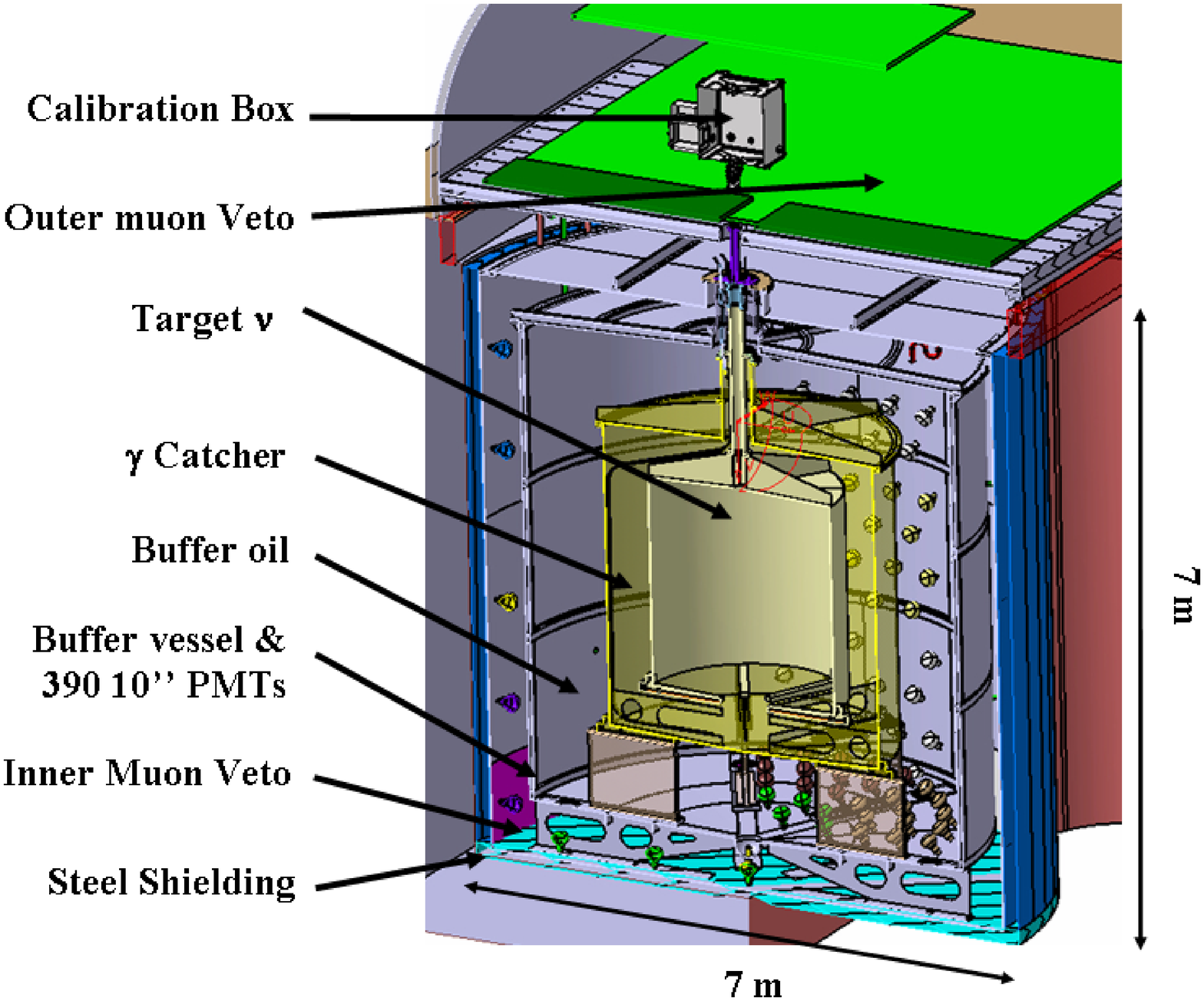}
  \caption{The Double Chooz detector.}
 \label{fig-DC}
\efi

\section{Effects of the magnetic field on the Hamamatsu R7081 10'' PMTs}
\label{pmt_resp}

\subsection{The Hamamatsu R7081 10'' photomultiplier}

The PMT Hamamatsu R7081~\cite{r7081} of 10 in. diameter has been chosen due to the suitable size for Double Chooz detector scale. An appropriate photocathode coverage of 13\% is reached using 390 PMTs. 
In addition, the PMT provided by Hamamatsu has a very low radioactivity glass and is oil proof.
 
The height of the PMT is 300 mm and the total weight is about 2500 g.
The cathode sensitivity range goes from 300 to 600 nm, being the maximum quantum efficiency about 25\% at 400 nm. The dynode structure is {\em Box and Line} with 10 stages. The typical gain is about 10$^{7}$ for voltages about 1500 V.

Low magnetic fields, about 500 mG, are expected to modify the electron trajectory in our PMT.
As mentioned before, the magnetic field in Double Chooz detectors is expected to be always below 1 G and not uniform. 
A dedicated setup has been made to quantify the effect of B-fields up to 2 G on the PMT in the three perpendicular directions. 

\subsection{Experimental setup}

A device has been designed and built to provide a fixed and homogeneous magnetic field in any direction, compensating the Earth's field.
The device consists of three Helmholtz coils, each one placed perpendicularly with respect to each others, as it is shown in Fig.~\ref{fig-coils}, defining the coordinate system. The coils are large enough to provide a homogeneous field in a 500 x 500 x 500 mm$^{3}$ volume where a black box containing the PMT is placed. A current of 1 A feeding one Helmholtz coil yields a uniform field of 2 G in that direction.

As Fig.~\ref{fig-coils} shows, the PMT is vertically placed inside a light tight box which is long enough to accommodate also, on the top, the light source used for the tests. 
The PMT inside the black box is placed at the center of the Helmholtz coils system with the dynode chain symmetric to the Y axis. 

The relationship between current and magnetic (B) field is measured prior to the inclusion of the black box using a three-axis magneto-resistive sensor, HMR2300~\cite{sonda}, which measures in the range from -2 to 2 G with an accuracy of 1 mG. 

The light source is a blue LED (DHC-SLR030NB40) controlled by a pulser circuit designed at CIEMAT. The LED is placed at 70 cm from the PMT and its light is diffused to have a homogeneous illumination over the PMT photocathode. Studies performed on the LED circuit stability over time gives an RMS variation of 0.2\% and a maximum dispersion of 0.9\% over 13 hours.
A single RG-303 cable connects the PMT to the high voltage power supply and to the DAQ electronics. Therefore, a circuit is necessary to split the PMT signal from the high voltage: the so-called {\em splitter}, which is placed outside the black box. The signal cable is sent from the splitter to the ADC (CAEN V265) where the signal is integrated in a 250 ns window and digitalized, in such a way that 1 ADC count corresponds to 0.033 pC.
The pulser circuit and the QDC are VME standard boards and are controlled through a LabView acquisition program running on a PC. A diagram of the setup is shown in Fig.~\ref{fig-electr}.

\bfi
\includegraphics[height=10.cm,width=14.cm]{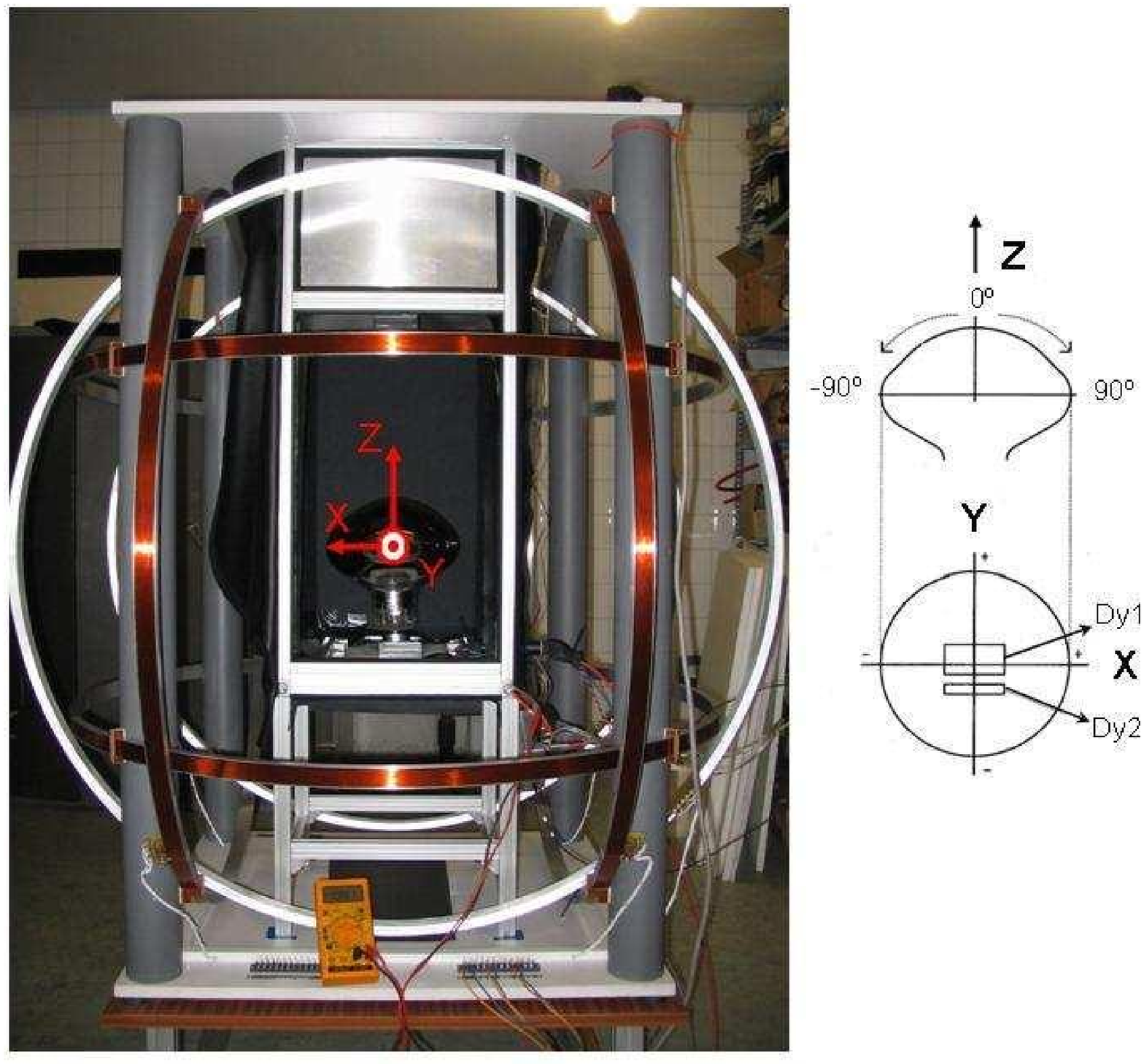}
  \caption{Helmholtz coils system. The light tight box is placed inside in such a way that the PMT is in the center of the system. The right plot shows the PMT position with respect to the Helmholtz coils: Dy1 is the first dynode and Dy2 is the second one, behind which the rest of the dynodes are placed.}
\label{fig-coils}
\efi

The measurement procedure is the following:
First, the Earth's magnetic field is compensated. The response of the PMT is quantified in this environment, as reference, and the LED is calibrated (pulse time vs. emitted light). Afterwards, the B-field is increased up to 1 G in 250 mG steps and from 1 G to 2 G in 500 mG steps in each of the 3 axis, keeping at zero the B-field in the other directions. At every step, the response of the PMT is analyzed

The expected effects of the B-field on the PMTs are:
\benum
\item A loss of collection efficiency due to the modification of the electron trajectory between the photocathode and the first dynode.
\item A decrease of the gain and a degradation of the single-photo-electron (spe) spectrum and the timing properties due to the deviation of electrons in the amplification chain.
\eenum
Disentangling both contributions would be only possible in spe regime by measuring the modified gain. 
However, a large amount of light would allow to better evaluate the global effect of the magnetic field up to higher fields than for the spe and would help to compute the effect over the collection efficiency.
Therefore, the PMT response is tested at two light levels: spe and about 50 photo-electrons (pe).

All the following measurements have been performed on two different R7081 PMTs showing very similar results. Accordingly, only the results of one of them will be reported in the next sections.

\bfi
\includegraphics[height=6.cm,width=12.cm]{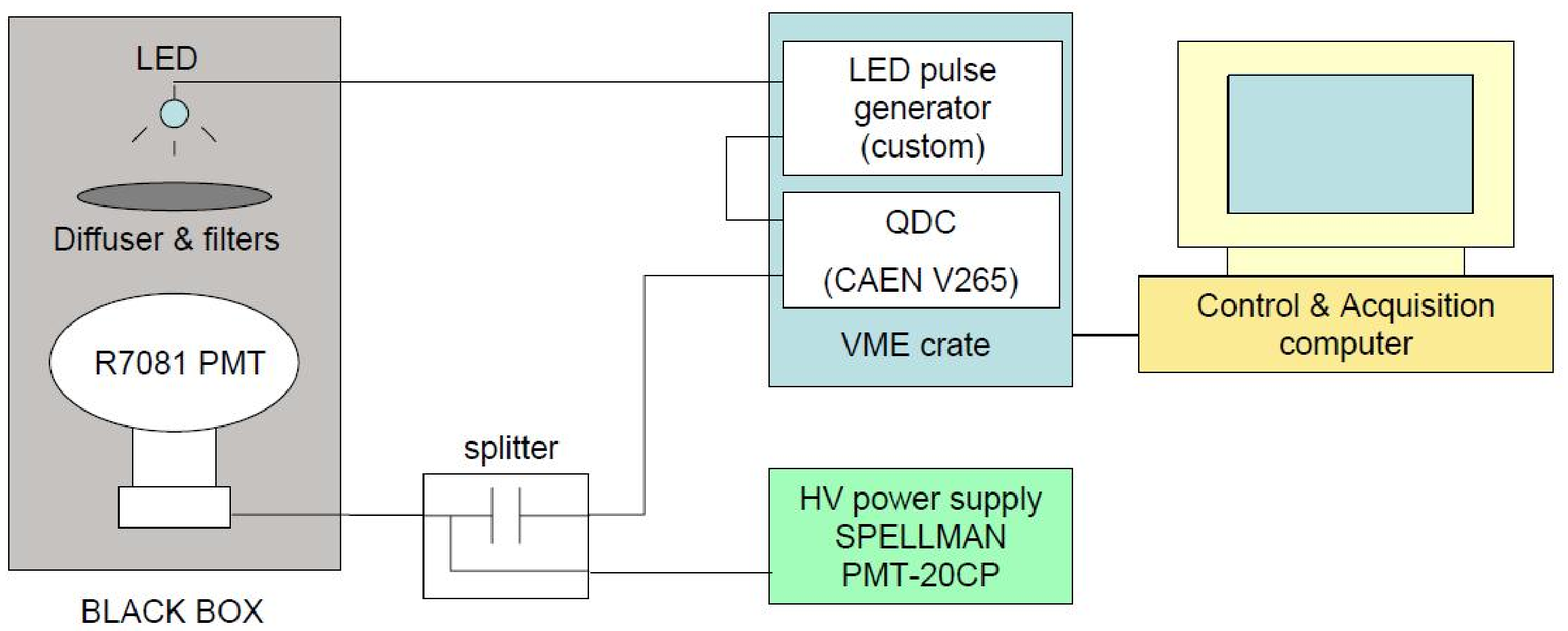}
  \caption{Diagram of the experimental setup}
\label{fig-electr}
\efi

\subsection{Calibration in absence of magnetic field}

Once the Earth's magnetic field is compensated, the PMT is illuminated with a very small amount of light to obtain the, so-called, spe spectrum. In Fig.~\ref{fig-spe}, the spe spectrum of our PMT is shown.  
The gain can be obtained very precisely from the fit of the spe spectrum to a parametric function describing the PMT response~\cite{note}. However, a statistical procedure has been chosen for this study in order to determine the gain independently of the spectrum shape, which is expected to be strongly modified by the magnetic field. This method is based on the Poissonian distribution of the number of pe and on the fact that the spectrum is divided in two contributions: noise (pedestal Gaussian) and signal (the rest of the distribution). The average number of pe ($\mu$) is obtained from the number of events contained in the pedestal Gaussian\footnote{The probability to have no pe is $P(0)={\rm e}^{-\mu}$}. 
And, then, the gain is determined as the mean of the distribution, subtracting the pedestal, divided by the average number of pe previously determined.
The gain is measured to be 50 ADC counts, equivalent to 10$^{7}$ electrons \footnote{The voltage applied to the PMTs is tuned to obtain a gain of 10$^{7}$ electrons.}.  

The LED is also calibrated to obtain the pulse time window corresponding to a determined average number of pe for the PMT under study. Since the detected light includes the PMT collection efficiency, which could be modified by the magnetic field, the light emitted by the LED should be measured at zero B-field. 
The spe spectrum is obtained for a LED pulse time of 11 ns, which corresponds to 0.2 pe at zero B-field. On the other hand, to obtain 50 pe a pulse time of 80 ns is applied.

\bfi
\includegraphics[height=7.cm,width=9.cm]{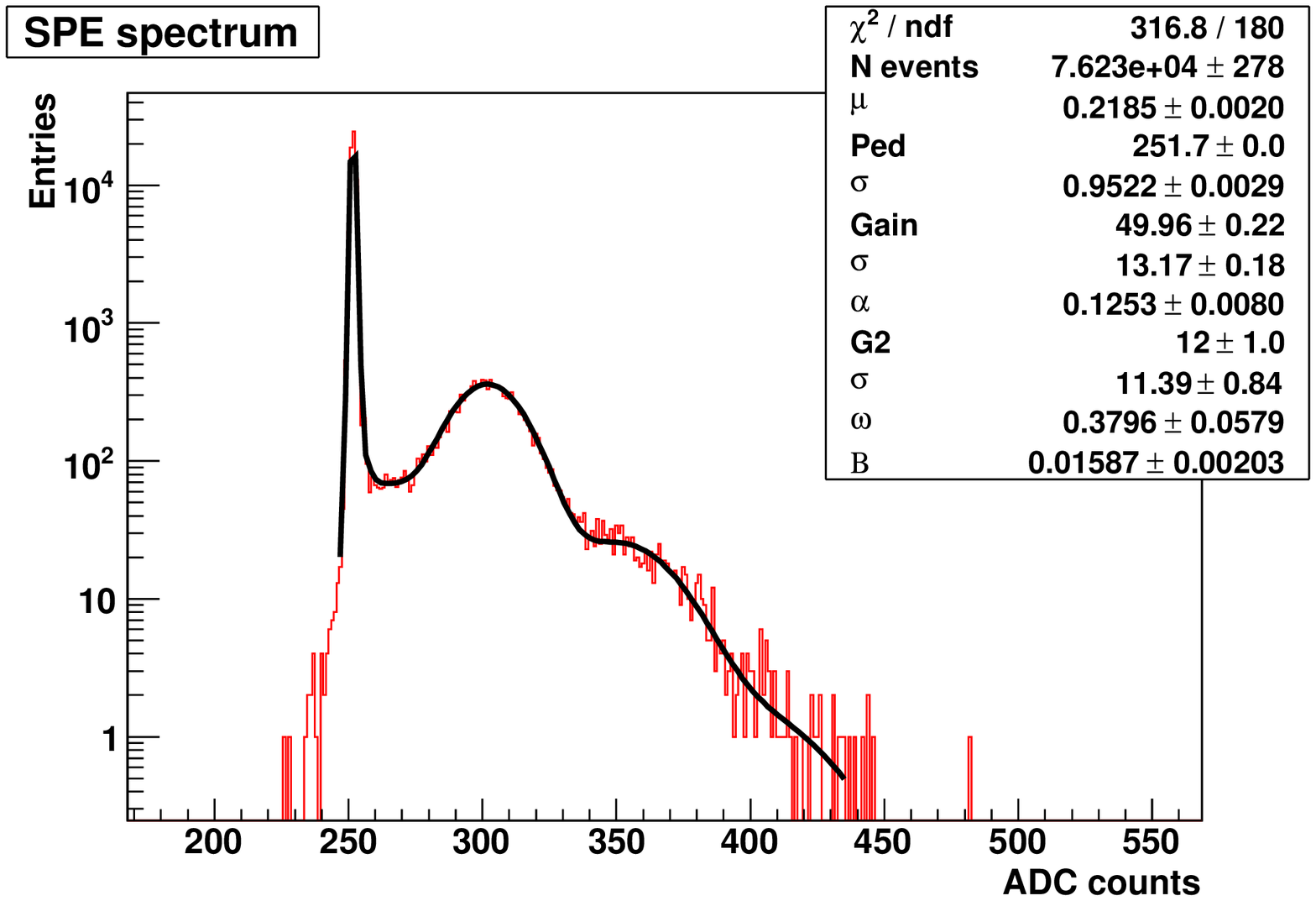}
  \caption{Spe spectrum from Hamamatsu R7081 PMT in absence of magnetic field. The function which parameterizes the PMT response is superimposed in black. The parameters describing the noise and signal are also shown~\cite{note}.}
 \label{fig-spe}
\efi

\subsection{Results for large amount of light}

In this regime, the PMT is illuminated with a LED pulse of 80 ns.
The PMT response for different values of the magnetic field along X,Y and Z directions is shown in Fig.~\ref{fig-bxyz}.
The largest effect is observed in -X direction (60\% of the signal is lost for 500 mG). As it is shown in Fig.~\ref{fig-coils}, a field in -X direction would drive the electron away from the first dynode. 
If the B-field is applied along the Z axis, the degradation of the PMT response is smaller because the Lorentz force is null for electrons moving vertically. On the other hand, the effect is symmetric in Y and Z, as expected.

A summary of these measurements is shown in Fig.~\ref{fig-1bxyz}. Low fields of the order of the Earth's magnetic field ($\approx$ 500 mG) along -X axis degrade the PMT response to a 40\%. 
In addition, a B-field of 1 G applied in any direction of the PMT transverse plane reduces the PMT signal by more than 80\%.

 \bfi
\includegraphics[height=5.cm,width=7.cm]{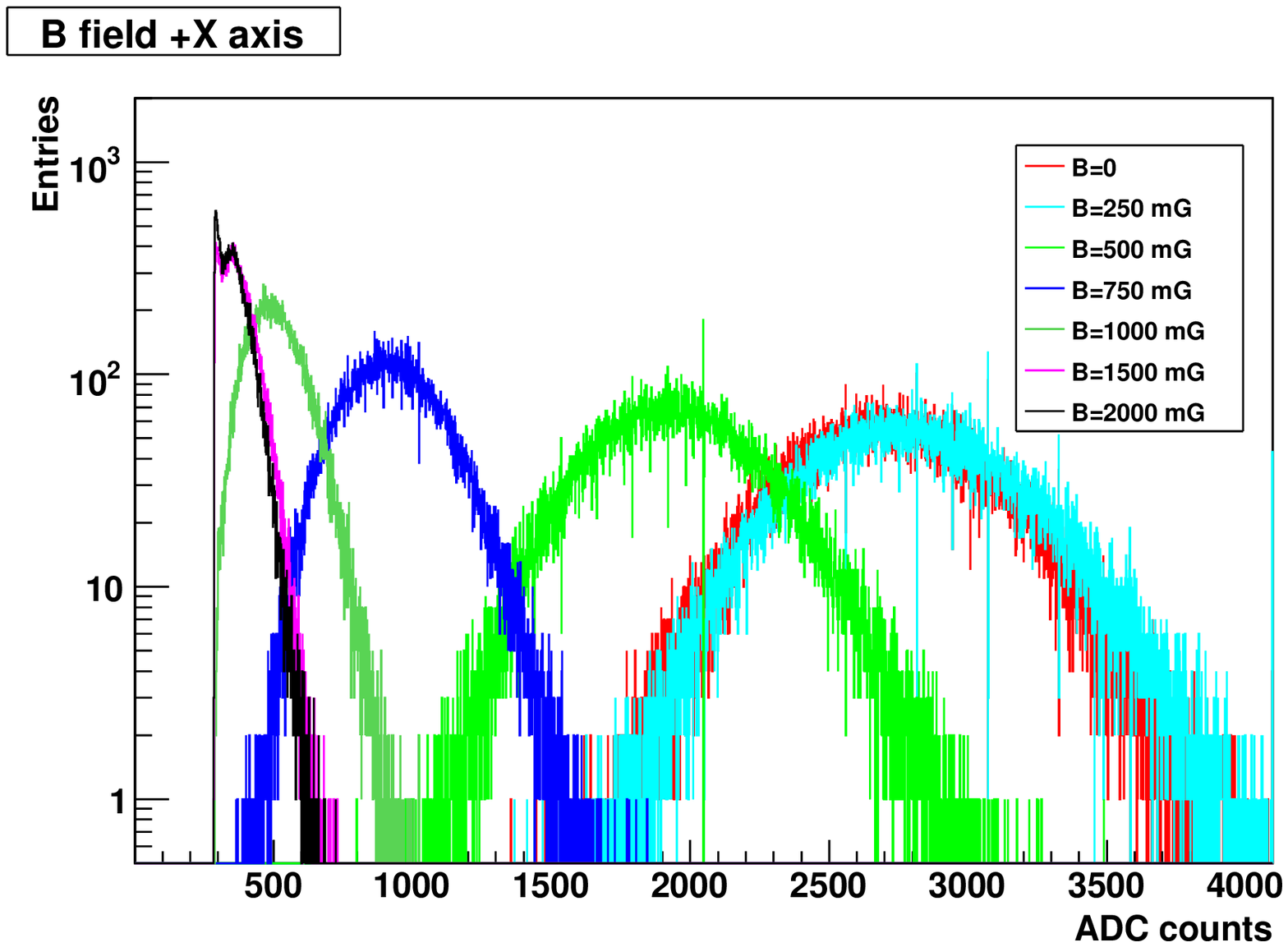}
\includegraphics[height=5.cm,width=7.cm]{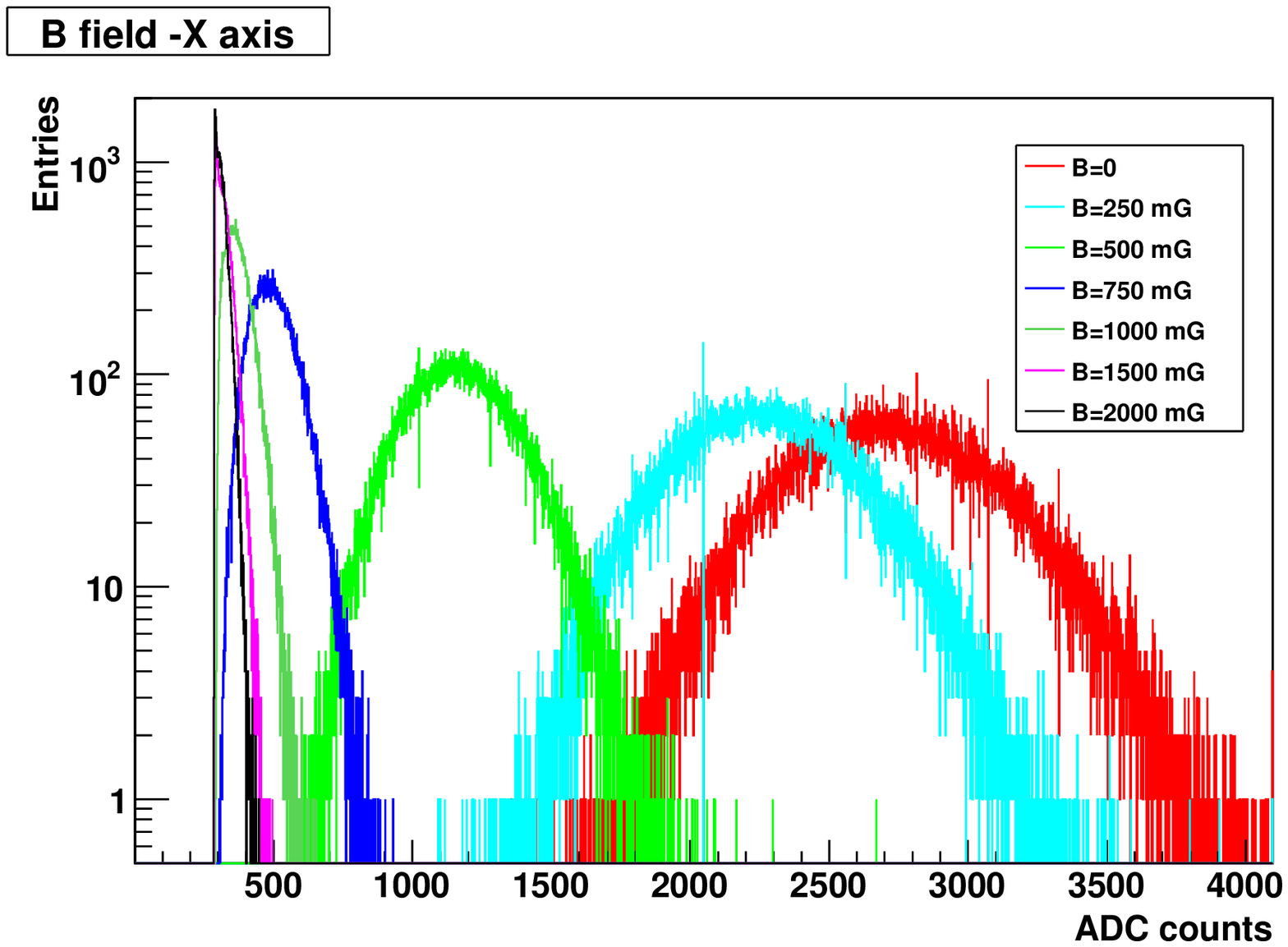}
\includegraphics[height=5.cm,width=7.cm]{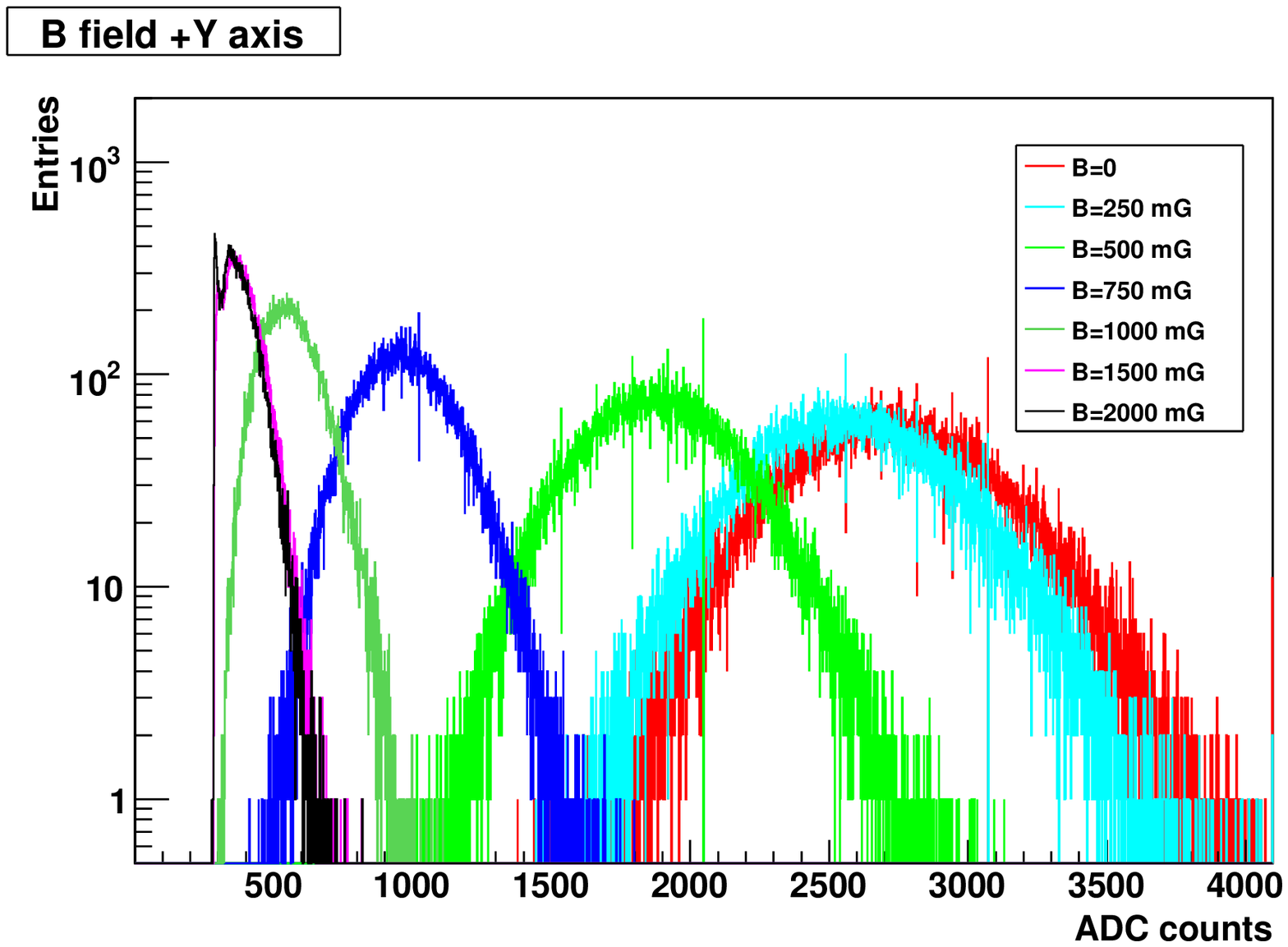}
\includegraphics[height=5.cm,width=7.cm]{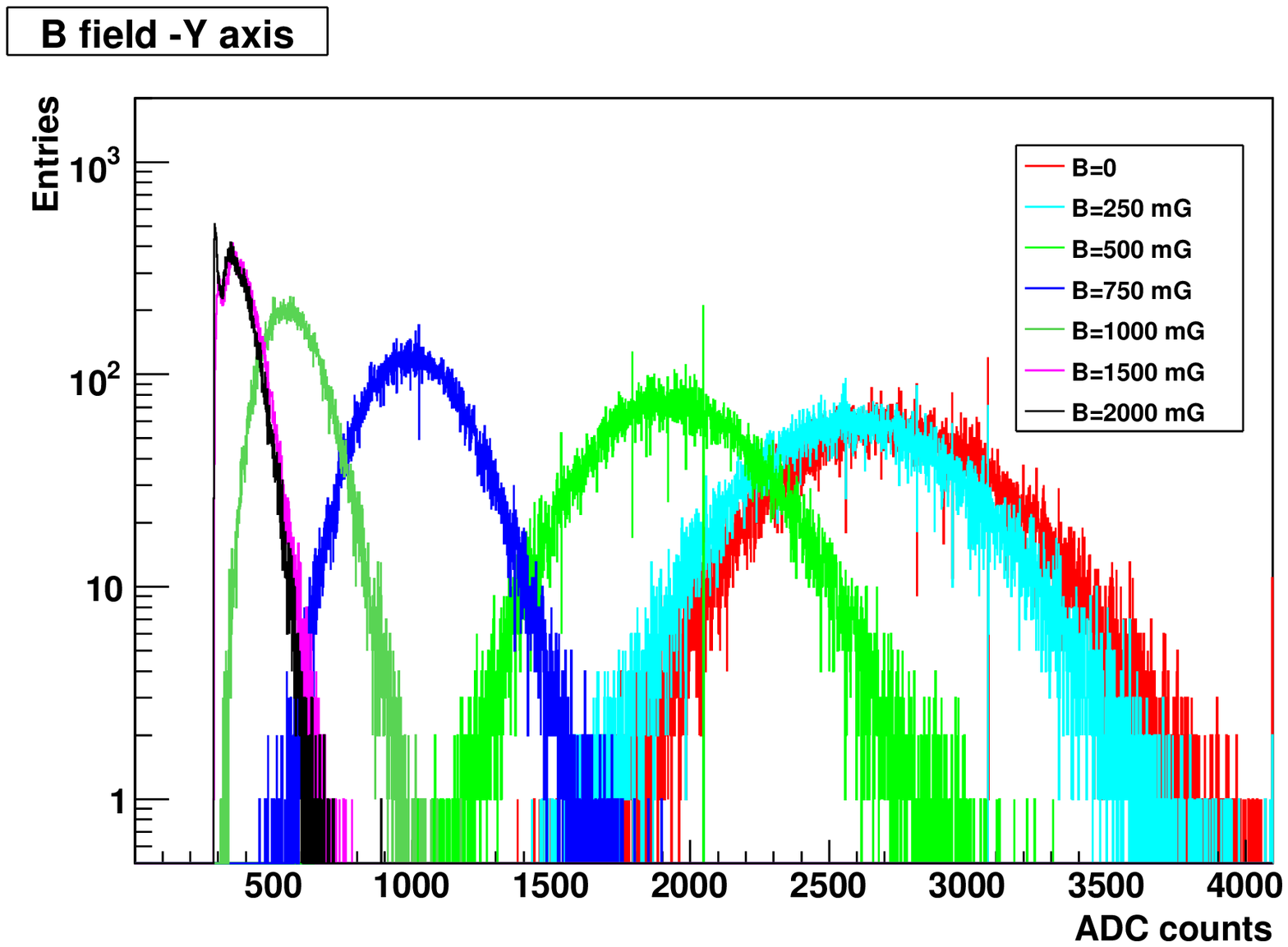}
\includegraphics[height=5.cm,width=7.cm]{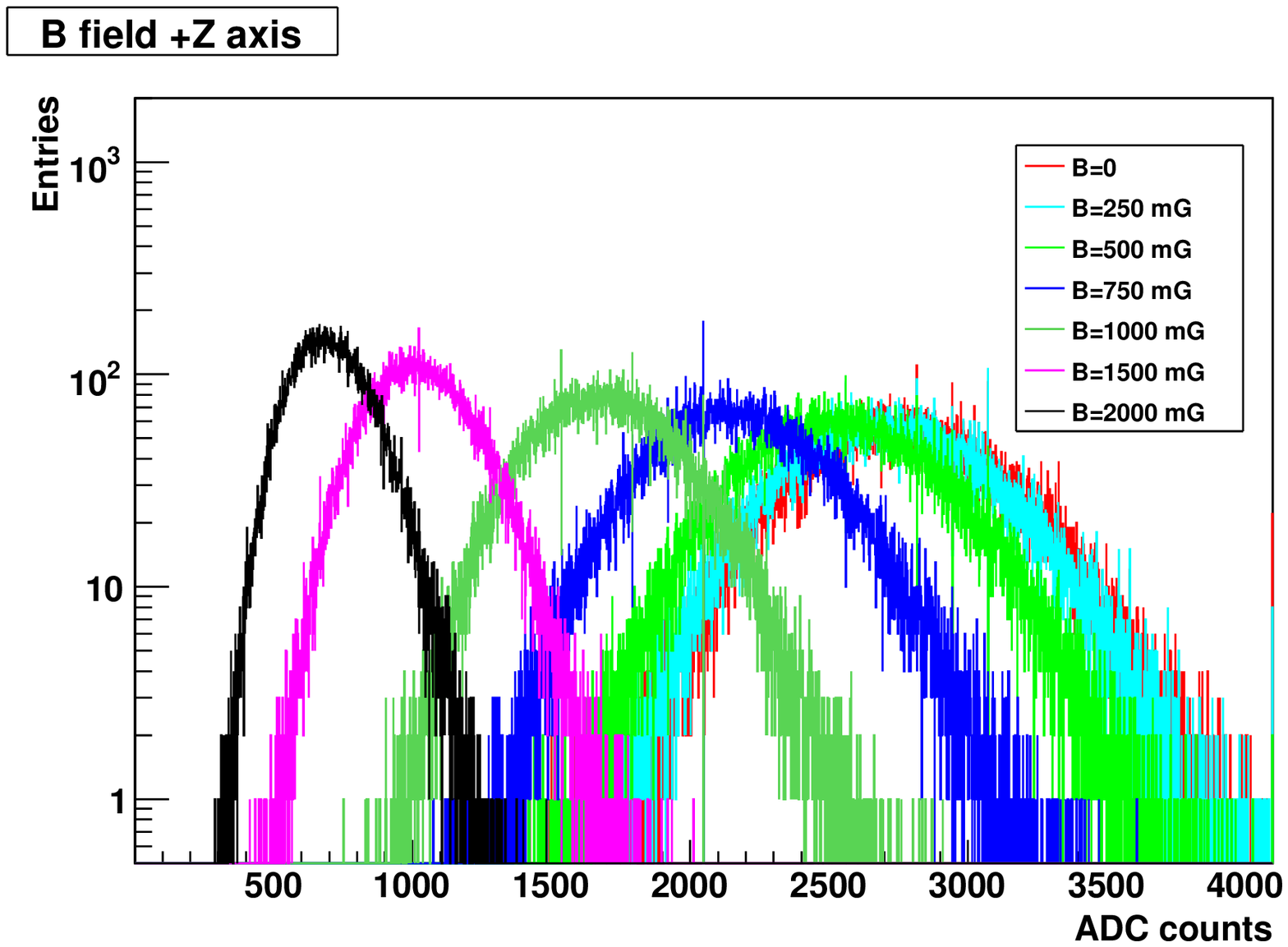}
\includegraphics[height=5.cm,width=7.cm]{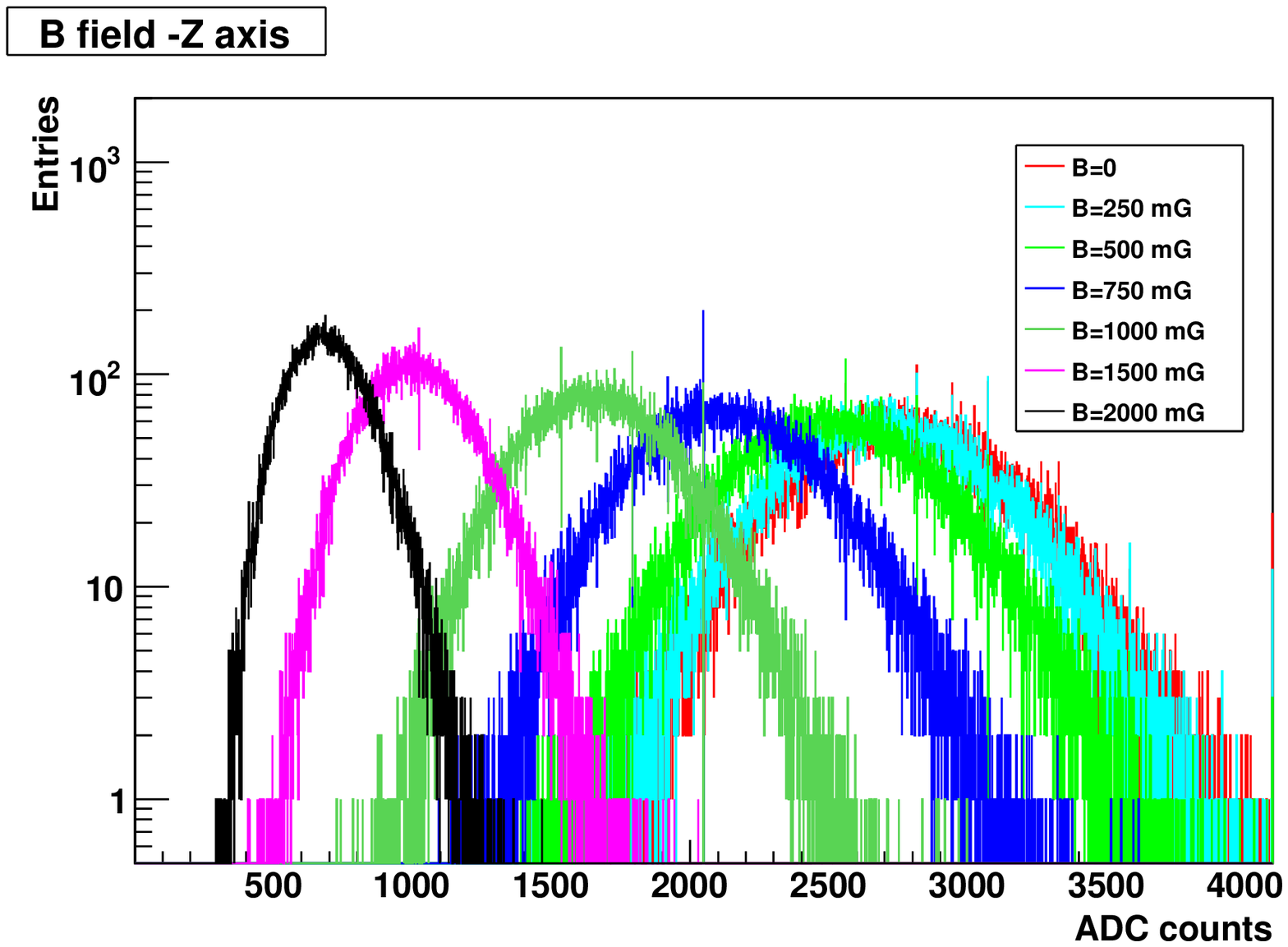}
  \caption{PMT response to a LED pulse time of 80 ns. The two top plots correspond to a B-field along X direction, the middle ones to a B-field along Y axis and the bottom ones to the Z direction.}
 \label{fig-bxyz}
\efi

\bfi
\includegraphics[height=7.cm,width=9.cm]{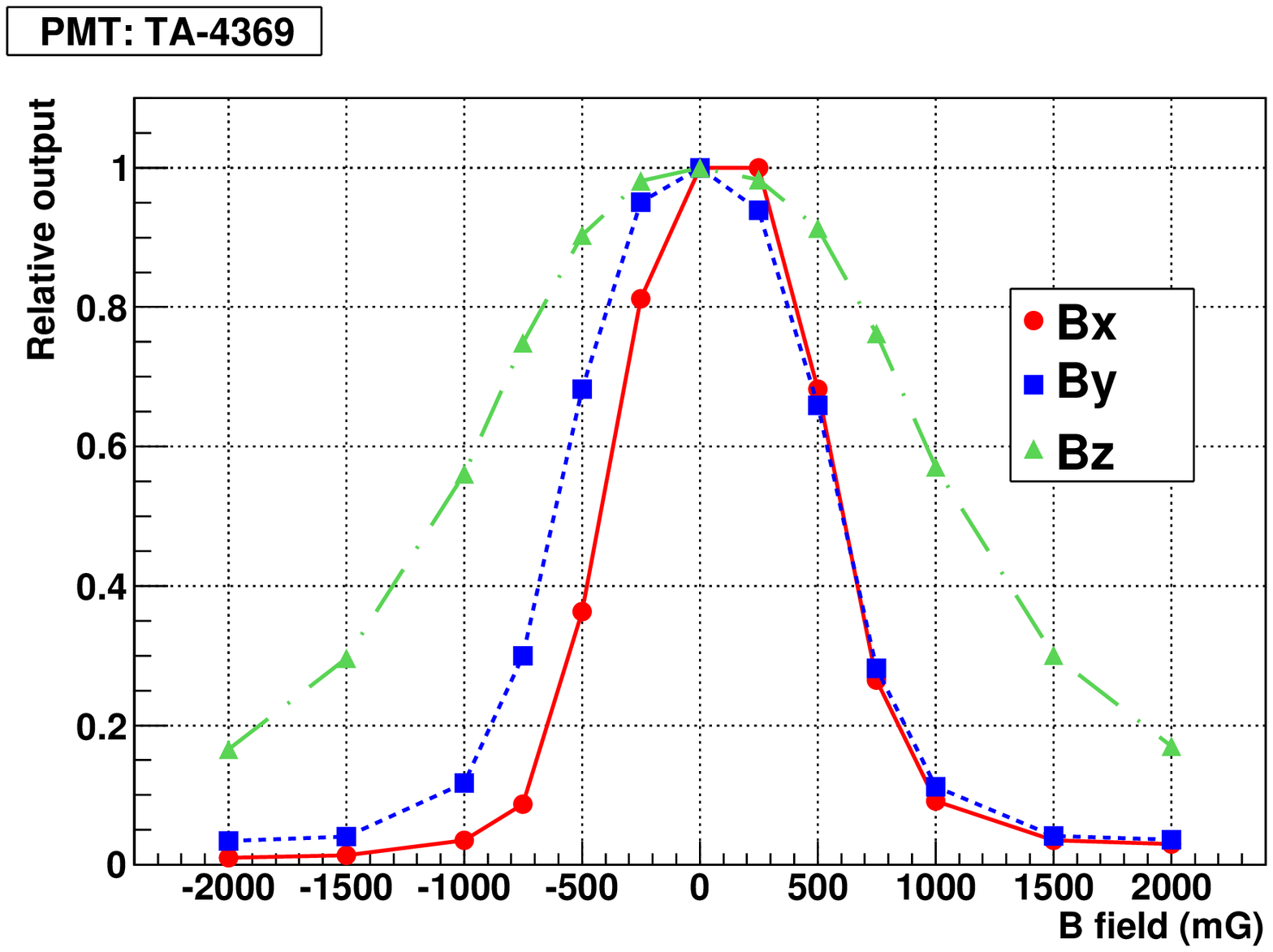}
  \caption{Relative response of the PMT with respect to zero B-field for different values of the magnetic field. The red curve corresponds to values of B along X direction, the blue one to Y axis and the green one to a B-field along Z axis.}
 \label{fig-1bxyz}
\efi

\subsection{Results for spe range}

In Double Chooz the study of the effect of the B-field on the spe spectrum signal is crucial because the expected number of pe per PMT for the neutrino signal is about 1.5.
In our setup, the PMT is illuminated with a light pulse of 11 ns.
Fig.~\ref{fig-spe_0_500} illustrates the effect of the magnetic field. The spe spectrum for zero B-field (black histogram) is plotted together with the same spectrum for a magnetic field of +500 mG along Y axis (red histogram). First, it is observed that the peak from the signal of one pe is moved to a lower value and the width of such a peak is larger due to the deviation of the electrons from their trajectory along the dynode chain. Second, the region between pedestal and signal (valley) is populated by badly amplified pe.
A way to quantify the degradation of the spe signal is through the measurement of the peak-to-valley (P/V) ratio. This magnitude takes into account the size of the peak (good amplified photo-electrons) and the population of the valley (very bad amplified pe). The measured P/V ratio for different B-fields is shown in Fig.~\ref{fig-pv}. A PMT with a P/V ratio below 2.5 does not fulfill the experiment requirements. This happens for B-fields about 500 mG applied perpendicularly to the PMT.     

In addition, the spe spectrum allows to analyze separately the effect of the B-field over the PMT gain and over the collection efficiency.
The effective gain and the average number of pe ($\mu$) are recalculated for B-fields up to 1 G. 
The relative variation of the collection efficiency is equivalent to the variation of $\mu$ if the light source is constant. The variation of $\mu$ is obtained from the strong illumination regime because the LED in this last case is less sensitive to temperature and voltage supply variations than in the spe range.

The relative variation of gain and collection efficiency as a function of the applied B-field are shown in Fig.~\ref{fig-gain-eff}.  
The effect of the magnetic field is larger on the collection efficiency than on the gain, as expected for these large photocathode area PMTs. The asymmetric behavior of the PMT gain for B-fields applied along X direction is very important due to the asymmetry of the dynode chain in this axis. The gain even increases for positive fields because the photo-electrons are more effectively focused from first to second dynode.  

\bfi
\includegraphics[height=7.cm,width=9.cm]{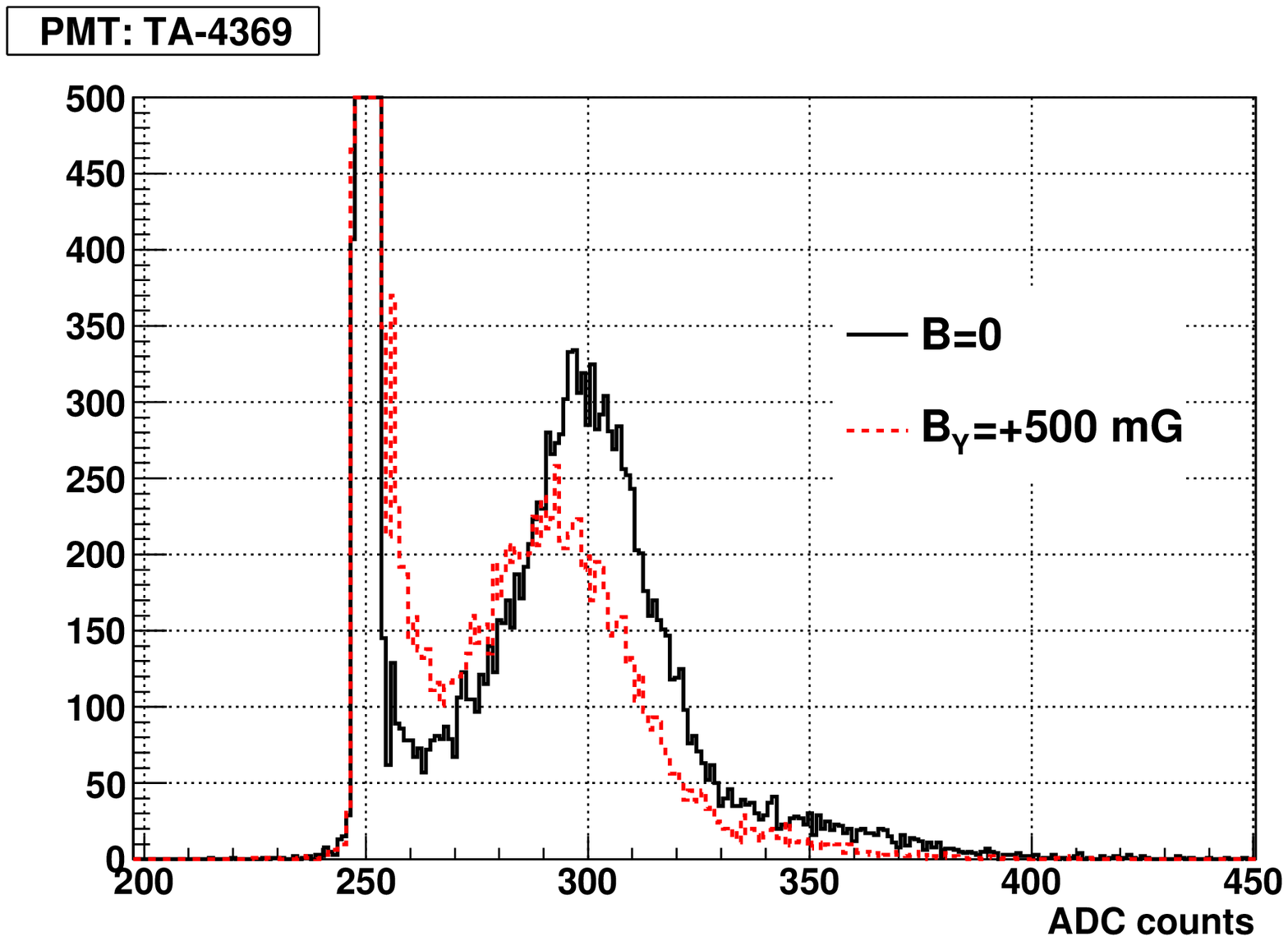}
  \caption{Spe spectra of the PMT illuminated by a LED pulse of 11 ns. The black histogram corresponds to the response of the PMT for B=0 and the red one for a B-field of 500 mG along Y direction.}
 \label{fig-spe_0_500}
\efi

\bfi
\includegraphics[height=7.cm,width=9.cm]{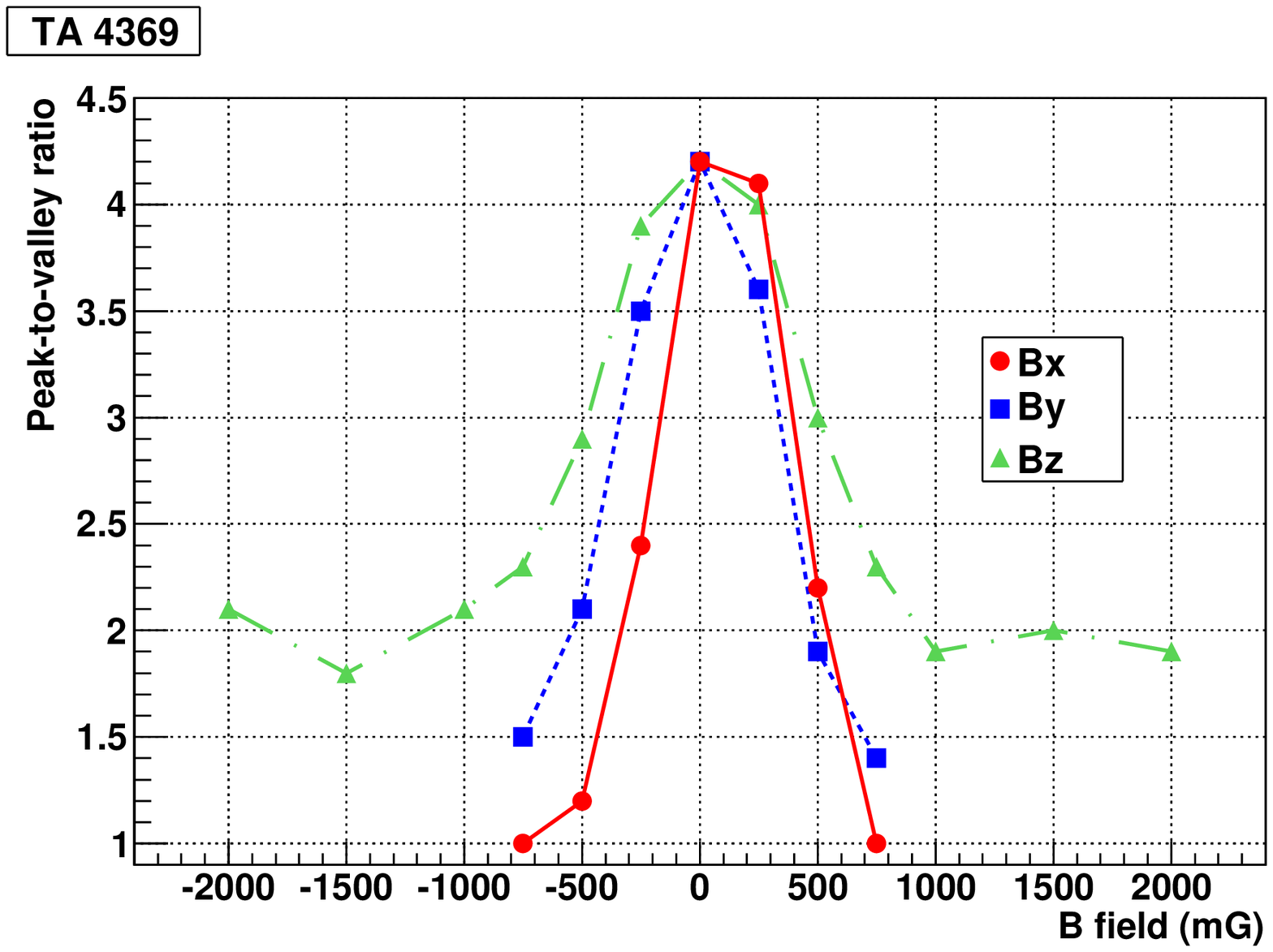}
  \caption{Peak-to-valley ratio as a function of the applied magnetic field. The red curve corresponds to values of B along X direction, the blue one to Y axis and the green one to a B-field along Z axis.}
 \label{fig-pv}
\efi

\bfi
\includegraphics[height=5.cm,width=7.cm]{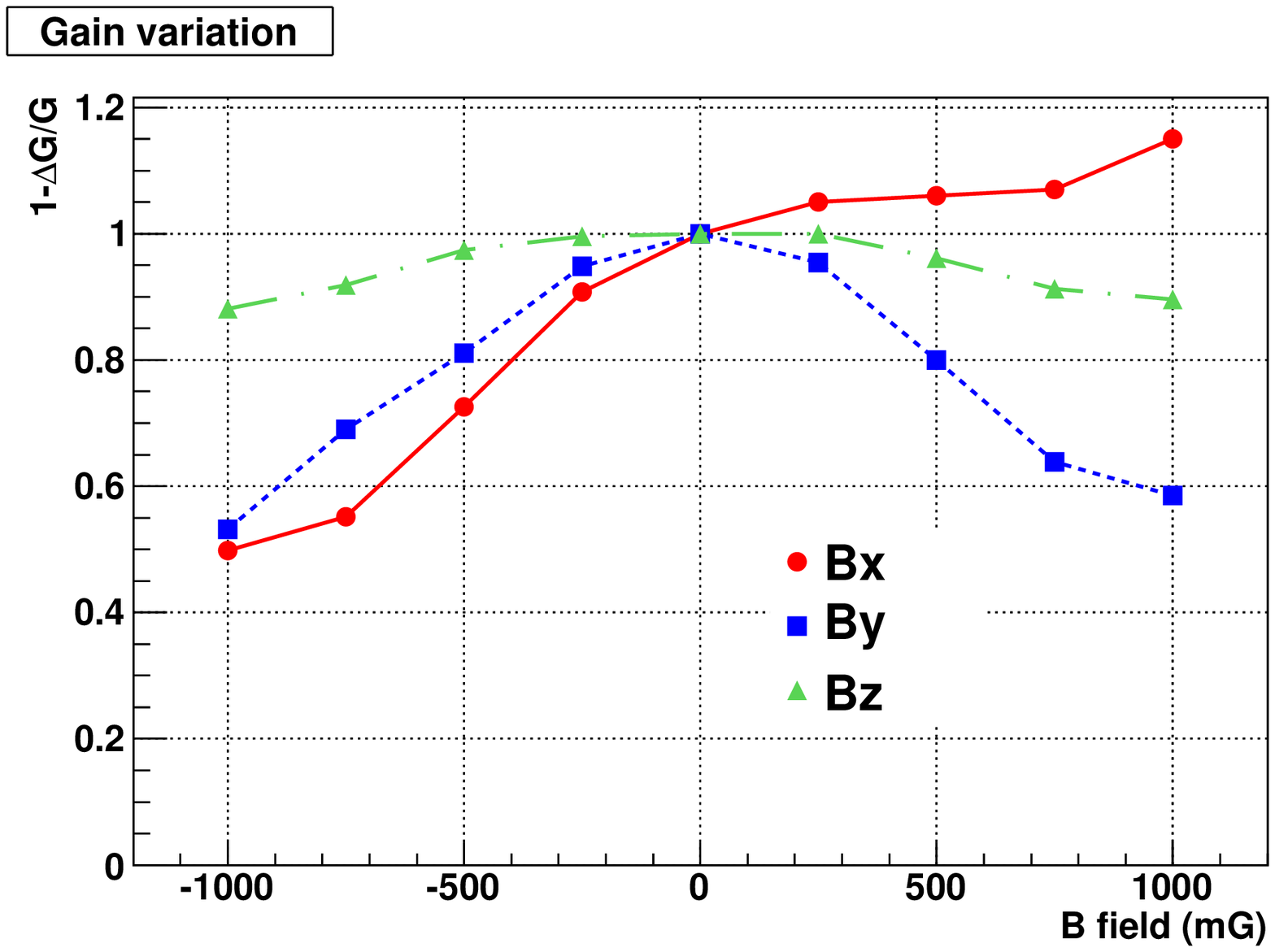}
\includegraphics[height=5.cm,width=7.cm]{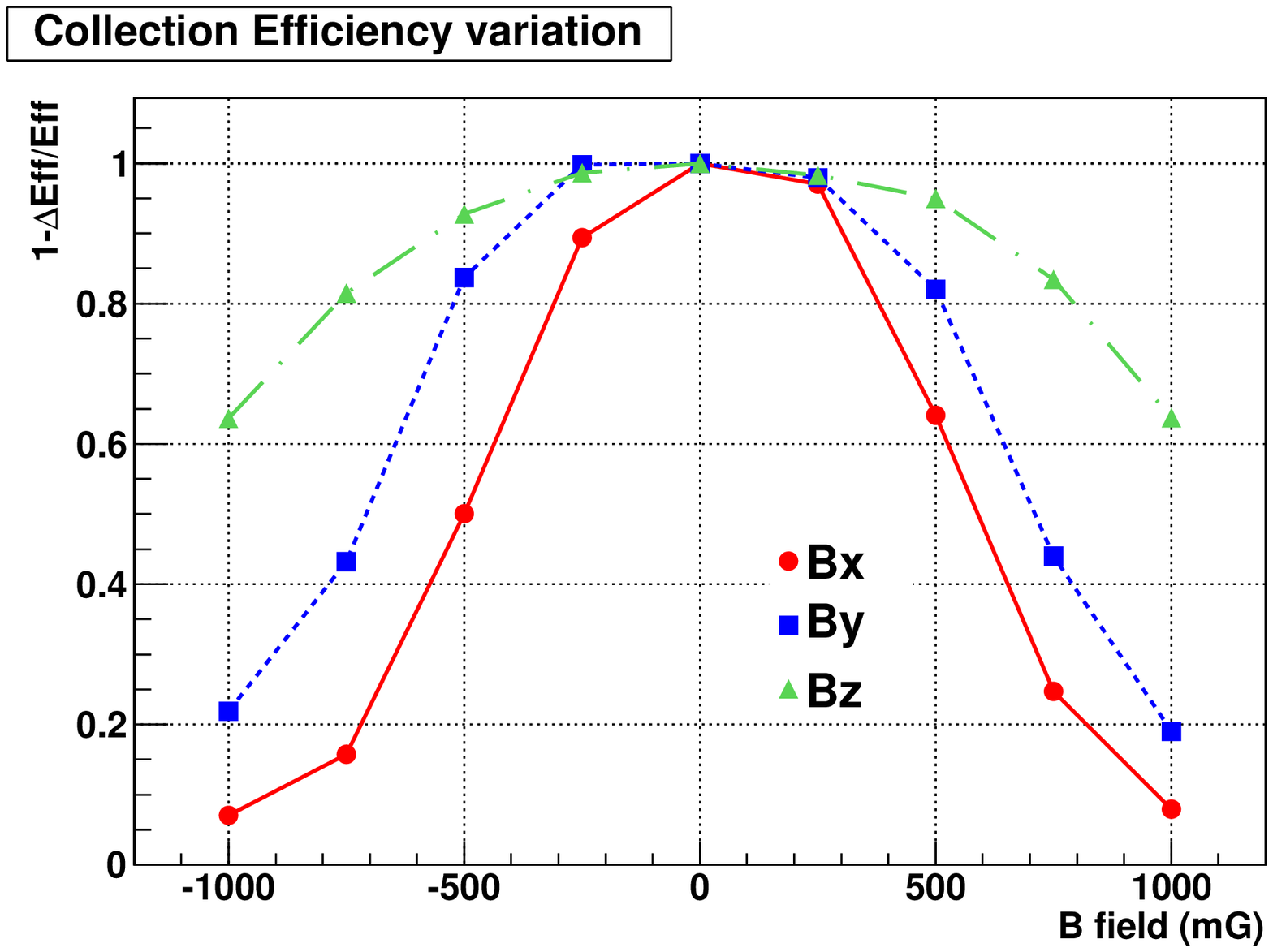}
  \caption{Variation of the gain (left) and collection efficiency (right) as a function of the applied B-field.}
 \label{fig-gain-eff}
\efi

\subsection{Discussion of the results}

The study presented in this paper shows that the Hamamatsu R7081 PMT response is reduced between 30\% and 95\% for transverse B-fields from 500 mG to 1 G. This so large effect will reduce dramatically the performance of our detectors. 
A degradation of the PMT response will affect the measurement of the energy due to the loss of collected pe. A worse energy resolution would affect the separation signal to noise and the sensitivity of the neutrino energy spectrum to $\theta_{13}$. On the other hand, since the B-field is expected to be 
different in both detectors, an accurate PMT calibration should be required in order to avoid additional uncertainties between them. 
  
We have observed that the gain can be recovered increasing the applied voltage but not the collection efficiency.
Therefore, a magnetic shielding becomes mandatory to keep the good performance of these PMTs. 

\section{Design and optimization of the PMT magnetic shields for the Double Chooz experiment}
\label{shield}

A possible way to shield the PMTs against the magnetic field is by using compensating coils for the whole detector. However, this solution presents several problems: It does not cancel completely a non-uniform magnetic field and uses an extra space not available in the tight pit of the Chooz lab.
Therefore, the Double Chooz collaboration has chosen a passive individual shield for every PMT.

\subsection{Design}

CIEMAT group has designed a passive magnetic shield considering the different constraints and requirements. 
The loss of PMT signal due to the presence of the magnetic field is required to be smaller than 10\%. Considering the results reported in the previous sections, that is equivalent to accept a transverse B-field smaller than 150 mG.
The performance of the shield is measured by means of the transverse shielding factor, defined as the ratio between the external and internal transverse B-fields. If the maximum external field is 2 G (a conservative value for the design), the required shielding factor should be 13. 

The simplest design is a cylinder without endcaps fitting the PMT dimensions: 300 mm diameter and about 300 mm high. 
The material and its thickness were chosen according to its permeability. 
The mu-metal\footnote{Mu-metal is a registered trademark of Magnetic Shield Corp.~\cite{magsh}.} is a very known high permeability material that satisfies our requirements. A very detailed study~\cite{nim_shield} has shown that the magnetic properties of the material as specified by the manufacturer are not reliable at these so very low fields. In the gauss range, the magnetic behavior of the mu-metal depends strongly on the size of the magnetic domains, which depends, in turn, on the mu-metal sheet thickness and thermal treatment, and should be determined experimentally for each sample. 
Several materials of different thickness and heights have been tested and some of them satisfied our requirements. Finally, the mu-metal from Meca Magnetic Company~\cite{mecam} of 0.5 mm thickness was chosen due to its very good performance and other practical aspects as the price and delivery time. The length of the shield is reduced to 275 mm because of mechanical constraints. 

The last step in the shielding design is to define the position of the PMT inside the cylinder.  

\subsection{Study of the PMT positioning inside the shield}

The transverse shielding factor is not homogeneous along the mu-metal cylinder axis. The shielding factor (SF) is maximum at the center and decreases steeply as moving away from the center because of the openings. Fig.~\ref{fig-sf} shows an example of this behavior. In the openings, the SF does not depend on the material but only on the geometry of the shield.
The photocathode to the first dynode region is the most affected by the magnetic field and has to be shielded as much as possible. 
Therefore, the PMT should be placed below the shield edge, where the shielding factor is smaller than our requirements (SF$\approx 4$ for a 300 mm diameter cylinder). In addition, the PMT acceptance would be reduced due to the cylinder shadow over the photocathode area. The final shield position should be a compromise between the SF achieved and the acceptance of photons. 

\bfi
\includegraphics[height=7.cm,width=9.cm]{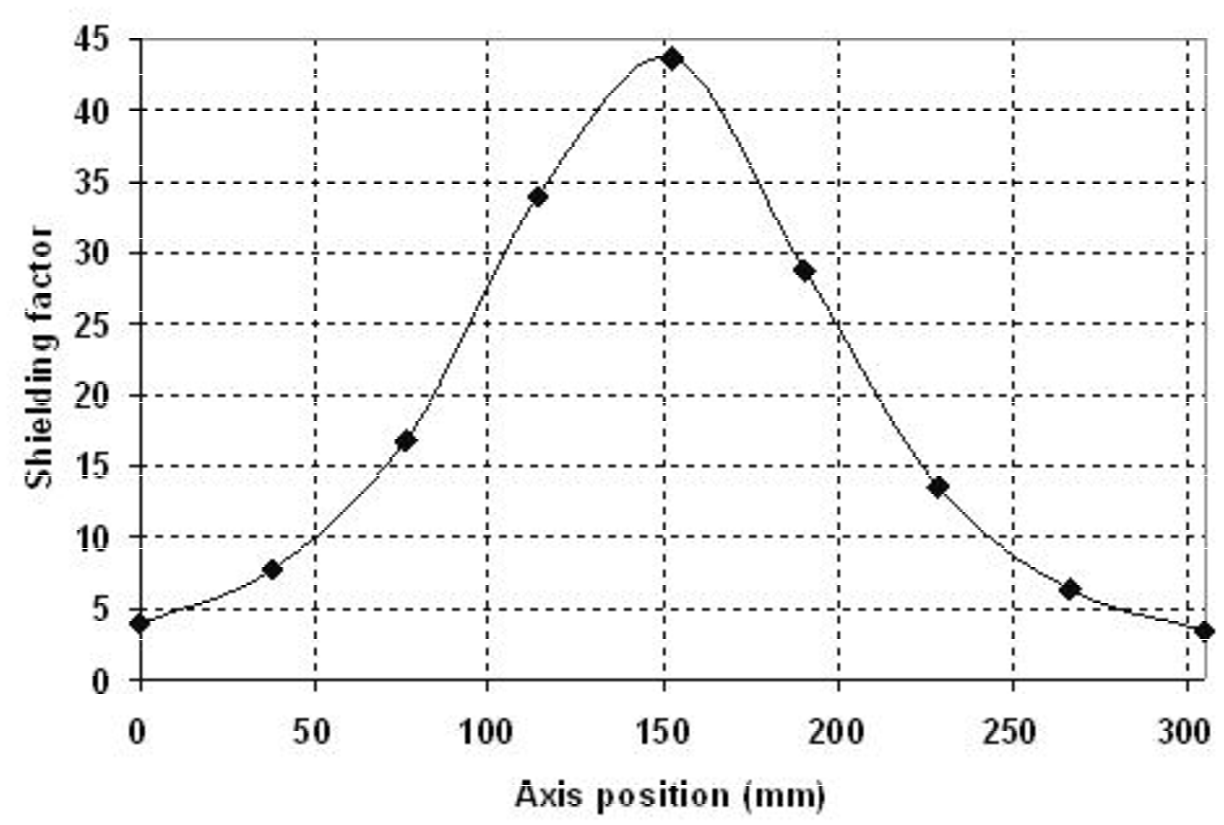}
  \caption{Shielding factor as a function of the position on the axis for a mu-metal cylinder from Meca Magnetic of 305 mm high, 280 mm diameter and 0.5 mm thick
in presence of a 2 G external field~\cite{nim_shield}.}
 \label{fig-sf}
\efi

The response of the shielded PMT is measured in the experimental setup shown in section 2.2 for different positions of the PMT inside the shield. This study has been done with a 0.25 mm thick cylinder of Magnetic Shield Corp..
The length of this cylinder is 380 mm and is moved vertically, defining its position by the length of the shield protruding above the PMT top. 
The internal walls of the shield were covered with a black sheet to avoid the reflected light that could distort our measurements.
A layout of the PMT inside the mu-metal cylinder is shown in Fig.~\ref{fig-lay}. The measurements have been done for 5 positions: -8.5 cm (shield edge below the PMT top, Fig.~\ref{fig-lay}-right) , 0 (at the PMT top level, Fig.~\ref{fig-lay}-center), 2.5 cm, 5 cm and 7.5 cm (shield edge above the PMT top, Fig.~\ref{fig-lay}-left).
The PMT performance has been tested for a large amount of light, applying the magnetic field separately along each axis while keeping null the B-field along the other coordinates.

\bfi
\includegraphics[height=4.cm,width=10.cm]{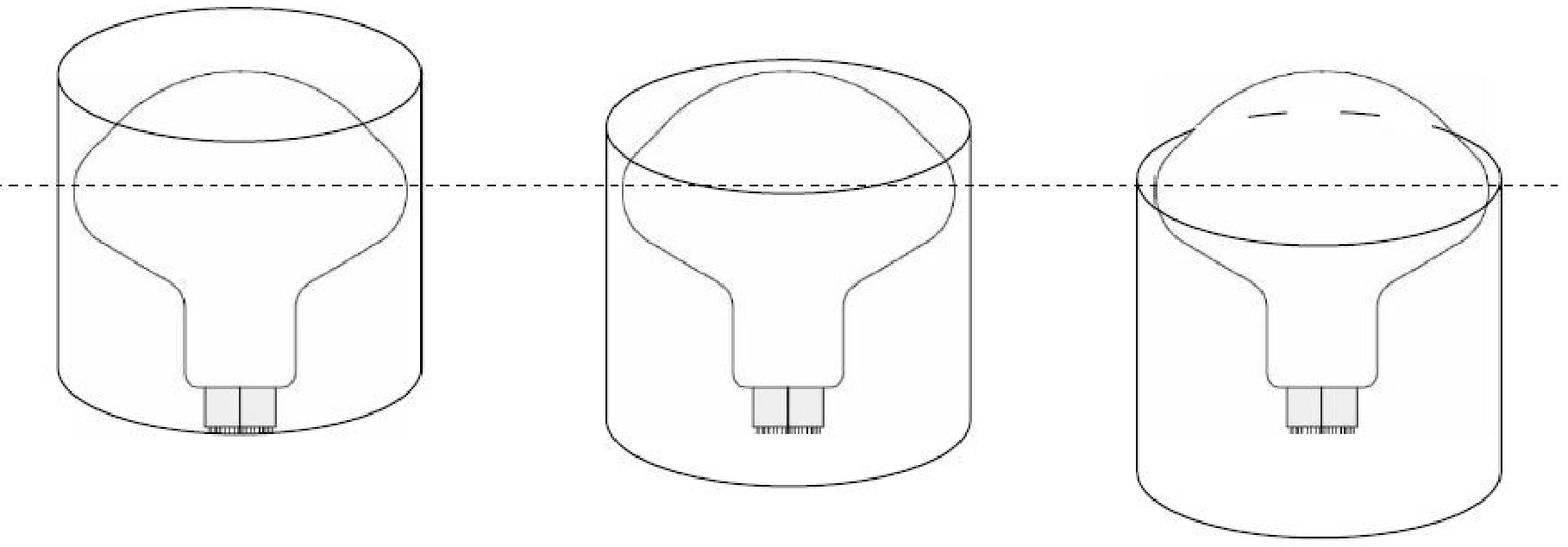}
  \caption{Layout of the PMT inside the mu-metal cylinder for three different relative positions.}
 \label{fig-lay}
\efi

The results are presented in Fig.~\ref{sh_pmt_pos_45}. The shielding is maximum in the transverse coordinates (X and Y), as expected, and satisfies our requirements even with the shield edge at the PMT top level (h$=$0). On the contrary, the vertical field (Z), for which smaller shielding is required, is very badly shielded due to the openings and depends much on the shield position. Moreover, for these vertical fields, a degradation of the PMT response with respect to the non-shielded PMT is observed when the cylinder is below the photocathode (h=-8.5 cm). The reason is that the lines of magnetic field are disturbed by the presence of the mu-metal and a transverse field is created in the PMT.
To summarize, the cylinder edge should be above the PMT top more than 2.5 cm to guarantee a degradation of the PMT response below 10\% up to 1 G. 

On the other hand, the cylinder has been included in the software simulation of the Double Chooz detector in order to determine the new acceptance for different heights above the PMT top. In Table~\ref{tab-sim}, the detected photon yield for an electron of 1 MeV generated at the detector center is shown for three positions of the PMT with respect to the cylinder. The result for a bare PMT is also shown as reference. The light detected by the 390 PMTs is reduced a 12\% when the shield edge is at the PMT top and does not decrease much when the cylinder protrudes from the PMT 5 cm or even 10 cm. 

\bfi
\includegraphics[height=5.cm,width=7.cm]{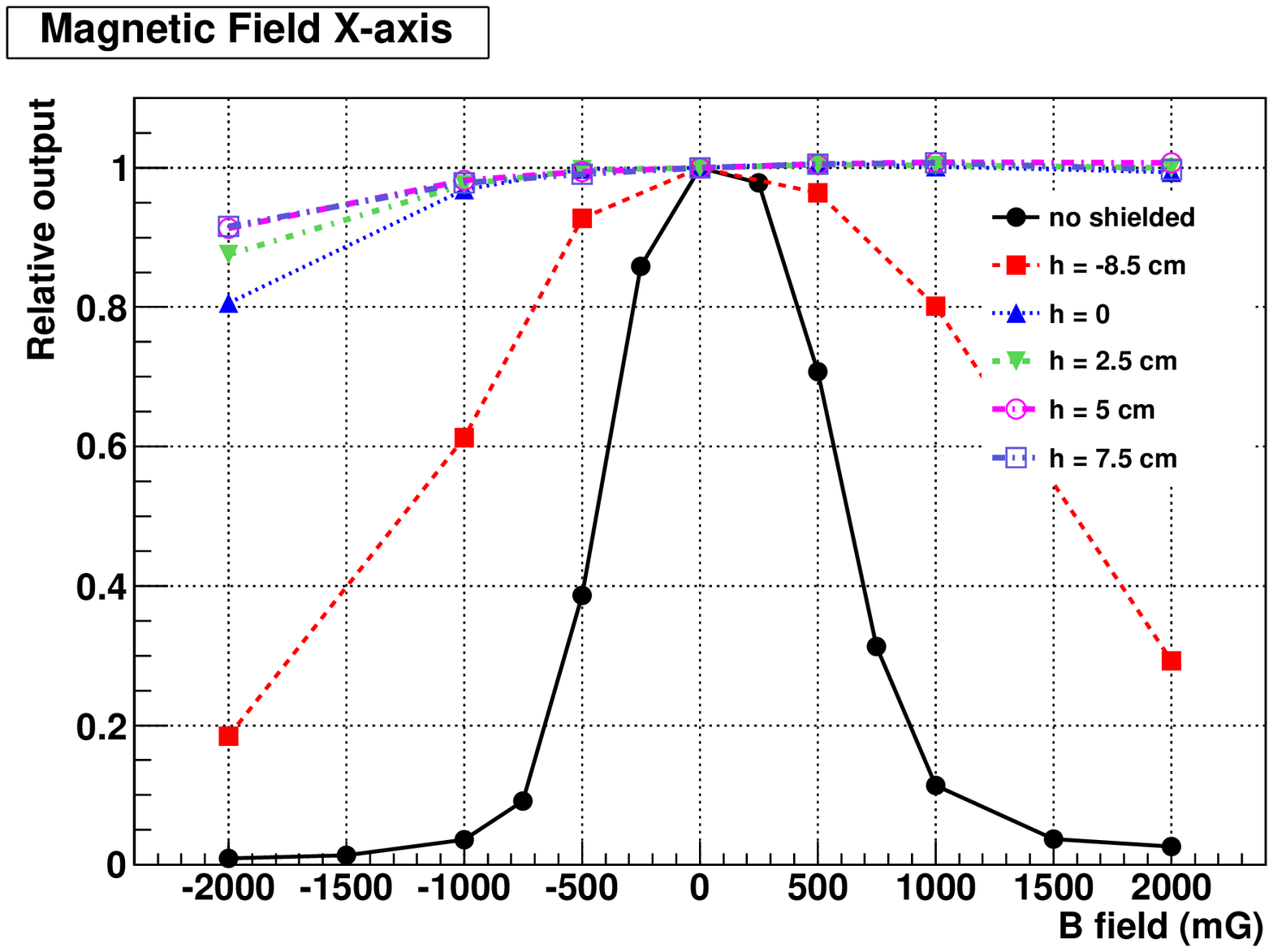}
\includegraphics[height=5.cm,width=7.cm]{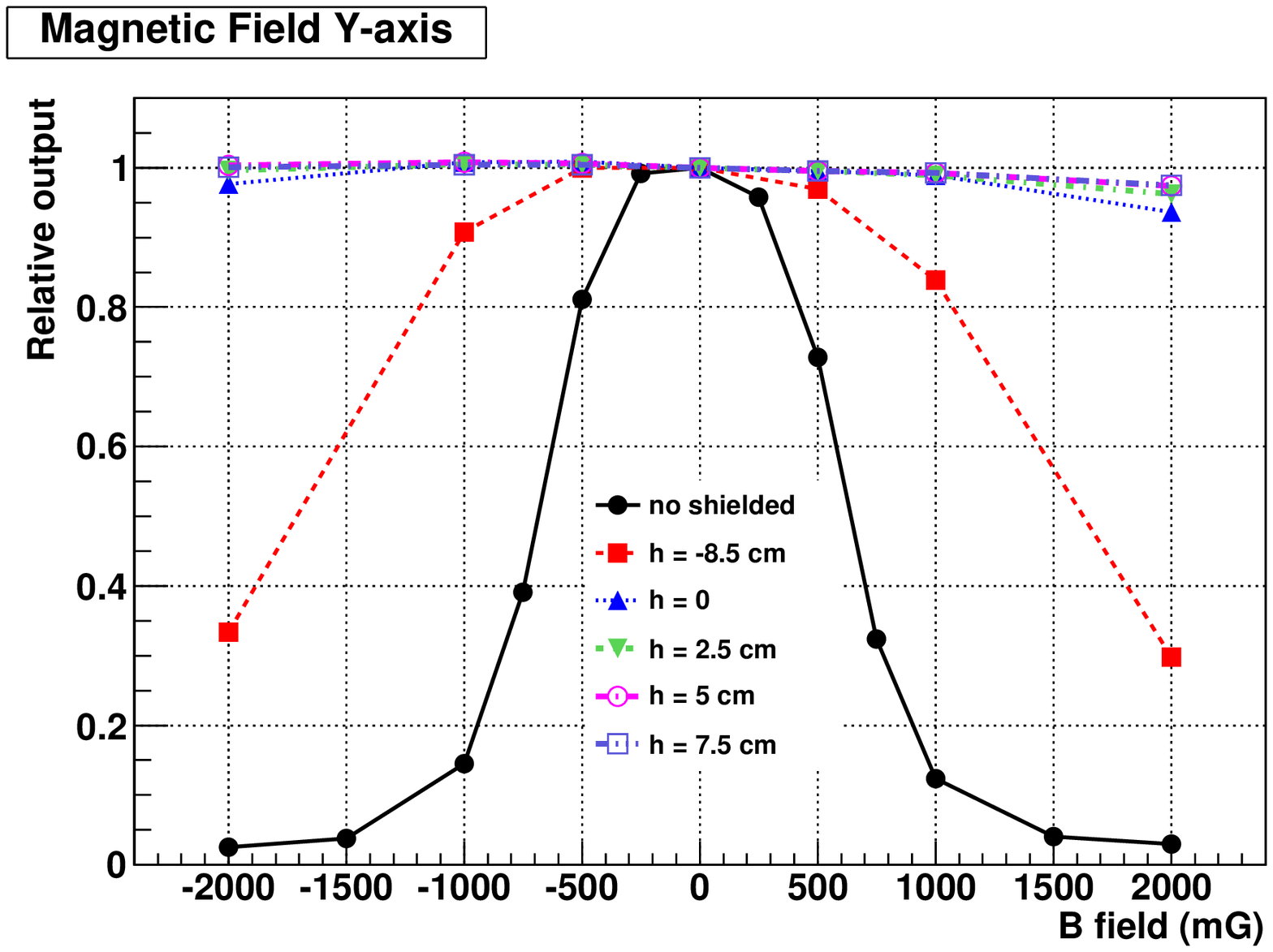}
\includegraphics[height=5.cm,width=7.cm]{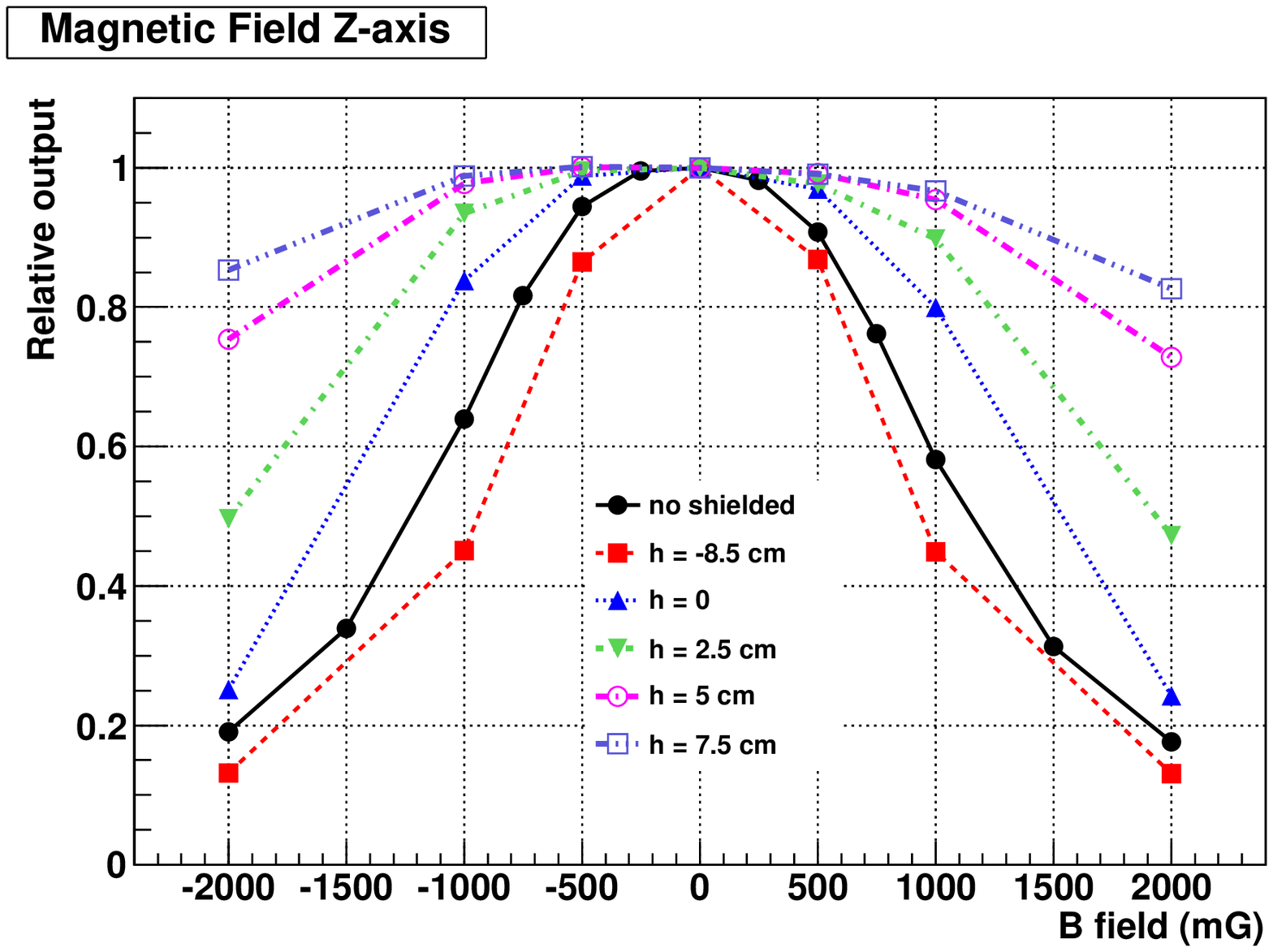}
  \caption{Relative output of the PMT with respect to zero B-field for different magnetic fields and positions of the mu-metal shield with respect to the PMT top (h). The black histogram corresponds to a non-shielded PMT.} 
 \label{sh_pmt_pos_45}
\efi

\begin{table}
\bc
\begin{tabular}{|c|c|c|}
\hline
H (cm) & Photon yield (pe) & Relative reduction (\%) \\
\hline
10 & 203.6 & 15 \\
5 & 205.2 & 14 \\
0 &  209.8 & 12 \\
\hline
Bare PMT & 239.3 & 0 \\
\hline
\end{tabular}
\ec
\caption{Photon yield measured by 390 PMTs for a 1 MeV electron generated at the Double Chooz detector center for different positions of the PMT inside the shield. H is the length of the shield above the PMT. The relative reduction on photon yield due to the shield shadow is also shown.}  
\label{tab-sim} 
\end{table}

A mu-metal cylinder from Meca Magnetic has also been tested in three different positions: 0, 2.5 cm and 5 cm, showing the same behavior but a slightly better performance (Fig.~\ref{fig-comp}). This result confirms our previous measurements in~\cite{nim_shield}. 

Taking into account that the acceptance of the PMT does not vary so much from 0 to 5 cm shield edge position, and considering also mechanical constraints for the PMT system support, the conservative solution of placing the shield 5.5 cm above the PMT top has been chosen.

\bfi
\includegraphics[height=5.cm,width=7.cm]{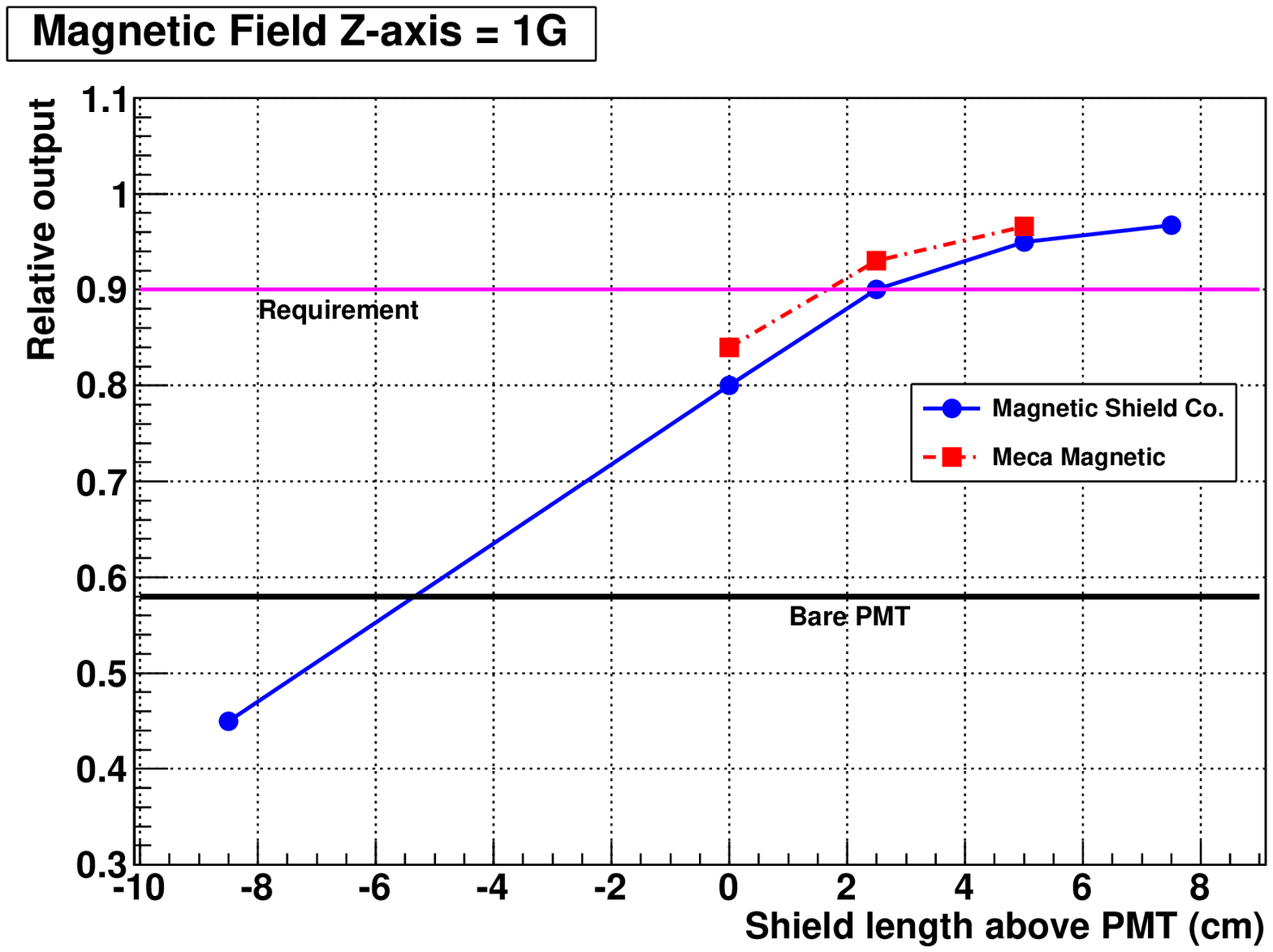}
  \caption{Relative output of the PMT with respect to zero B-field for ${\rm B}_{\rm Z}=-1$ G as a function of the shield length above the PMT. The result is shown for two of the materials under study.}
 \label{fig-comp}
\efi

\subsection{Performance of a shielded PMT under magnetic field}

Finally, the performance of the PMT has been tested with the shielding selected to be used in Double Chooz:
Meca Magnetic of 275 mm high, 300 mm diameter and 0.5 mm thick.
The distance between the top of the PMT and the shield edge was 5.5 cm.
Table~\ref{fin-1} presents  a summary of the performance of the shielded PMT under different B-fields. 
The possible degradation of the PMT response is measured through the signal of a large amount of light. The gain and collection efficiency were measured as mentioned in section 2.5. 
The errors (RMS) quoted in the same table have been determined propagating the statistical error of the corresponding variable.  
The effect of a magnetic field in Y is negligible up to 2 G. 
For magnetic fields along X axis, while the collection efficiency is fully recovered, the gain is still slightly affected. The most probable reason is that the lower edge of this shield is near the dynode chain and disturbs the lines of the magnetic field. As a result, an additional field along X is created distorting the amplification chain.
Finally, the effect of the B-field in Z is small enough, about 4\% at 1 G, to be controlled by calibration. 
To summarize, the loss of PMT signal for magnetic fields up to 1 G is always less than 5\% in any direction.

\begin{table}
\begin{tabular}{|c||c|c|c|c|c|c||c|}
\hline
${\rm B}_{\rm X}$ (mG) & -2000 & -1000 & -500 & +500 & +1000 & +2000 & RMS \\
\hline 
${\rm Signal}$(B$\neq$0)/${\rm Signal}$(B$=$0) (\%) & 90.7 & 95.4 & 97.5 & 102.2 & 104 & 107.3 & $\pm$ 0.15 \\
\hline
$\Delta({\rm Gain})/{\rm Gain}$  (\%) & -8 & -4 & -2.5  & +1.6 & +3.7 & +6.3 & $\pm$ 0.4 \\
\hline
$\Delta({\rm E}_{\rm coll})/{\rm E}_{\rm coll}$ (\%) & -1.3 & -0.6 & -0.04 & +0.6 & +0.2 & +0.9 & $\pm$ 0.5 \\ 
\hline \hline
${\rm B}_{\rm Y}$ (mG) & -2000 & -1000 & -500 & +500 & +1000 & +2000 & RMS \\
\hline 
${\rm Signal}$(B$\neq$0)/${\rm Signal}$(B$=$0) (\%) & 99.2 & 100 & 100 & 100 & 99.8 & 99.2 & $\pm$ 0.15 \\ 
\hline
$\Delta({\rm Gain})/{\rm Gain}$  (\%) & +0.1 & +0.9 & -0.4 & +0.5 & +0.3 & -0.3 & $\pm$ 0.4 \\
\hline
$\Delta({\rm E}_{\rm coll})/{\rm E}_{\rm coll}$ & -0.96 & -0.88 & +0.6 & -0.5 & -0.5 & -0.5 & $\pm$ 0.5 \\ 
\hline \hline
${\rm B}_{\rm Z}$ (mG) & -2000 & -1000 & -500 & +500 & +1000 & +2000 & RMS \\
\hline 
${\rm Signal}$(B$\neq$0)/${\rm Signal}$(B$=$0) (\%) & 72.7 & 96 & 99.2 & 99.6 & 96.7 & 75.2 & $\pm$ 0.15 \\
\hline
$\Delta({\rm Gain})/{\rm Gain}$  (\%) & -12.7 & -2.3 & -0.9 & +0.5 & -1.2 & -11.8 & $\pm$ 0.4 \\
\hline
$\Delta({\rm E}_{\rm coll})/{\rm E}_{\rm coll}$ & -16.7 & -1.7 & +0.18 & -0.9 & -2.2 & -14.8 & $\pm$ 0.5 \\
\hline
\end{tabular}
\caption{Response of the shielded PMT with respect to zero B-field for different values of B$_{\rm X}$, B$_{\rm Y}$ and B$_{\rm Z}$. The response of the PMT is described by its signal for a LED pulse time of 80 ns, its gain and the relative variation of its collection efficiency.}
\label{fin-1}
\end{table}

\section{Conclusions}
\label{conclu}

The measurements presented in this paper show that the PMT Hamamatsu R7081 is strongly affected by magnetic fields smaller than 1 G. The PMT signal is reduced more than 60\% for B-fields close to the Earth's magnetic field along X direction. A mu-metal cylinder wrapped around the PMT shields very effectively transverse magnetic fields, 
while the shielding in the vertical direction depends on the cylinder length protruding the PMT photocathode. 
It was demonstrated that, using a 275 mm high, 300 mm diameter and 0.5 mm thick cylinder with its edge placed 5.5 cm above the photocathode, the PMT response is reduced less than 5\% for B-fields up to 1 G in any direction.   
The installation of these shields in all PMTs of the two Double Chooz detectors will, therefore, ensure that no systematic effects due to magnetic field will degrade the expected sensitivity reach of the experiment in the measurement of $\theta_{13}$. 

\section*{Acknowledgements}
The authors would like to thank the members of the Double Chooz collaboration for their comments and suggestions on this work and, specially, F. Suekane for providing us with the Hamamatsu PMTs used in this study.


\begin{thebibliography}{00}
\bibitem{clim} Th. Schwetz, New J. Phys. 10:113011, 2008. 
\bibitem{CHOOZ} M. Apollonio et al., Eur. Phys. J. C27, 331-374, 2003.
\bibitem{DC} F. Ardellier et al., Double Chooz Collaboration, {\em Double Chooz: A search for the neutrino mixing angle $\theta_{13}$} Preprint arXiv:hep-ex/0606025, 2006.
\bibitem{exp} Super-Kamiokande: Y. Fukuda et al., Nucl. Instrum. Meth. A501, 418 (2003). Borexino: G. Alimonti et al., Astropart. Phys.16:205-234,2002. Antares: J.A. Aguilar et al., Nucl. Instrum. Meth. A555, 132-141, 2005.
\bibitem{r7081} jp.hamamatsu.com/products/sensor\verb2-2etd/pd002/pd394/R7081/index\_en.html.
\bibitem{sonda} Honeeywell Sensor Products, Smart Digital Magnetometer HMR2300, 900013902-04 Rev. H. Solid State Electronics Center (800), pp. 323-8295. www.magneticsensors.com
\bibitem{note} C. Palomares {\em Parameterization of the response of the photomultiplier Hamamatsu R7081}. Double Chooz Internal note EDMS I-014384.
\bibitem{nim_shield} E. Calvo et al., Nucl. Instrum. Meth. A600, 560 (2009).
\bibitem{magsh} Magnetic Shield Corp., Bensenville, IL 60106, USA. www.magnetic-shield.com.
\bibitem{mecam} Meca Magnetic, 490 Rue de la Fontaine, 45200 Amilly, France. www.magnetic-shielding.com.
\end{thebibliography}
\end{document}